\documentclass[12pt]{article}
\usepackage{amssymb}
\usepackage{graphicx}
\usepackage{amsmath}

\def\e{{\rm e}}

\def\d{\partial}

\newcommand{\be}{\begin{equation}}
\newcommand{\ee}{\end{equation}}
\newcommand{\bea}{\begin{eqnarray}}
\newcommand{\eea}{\end{eqnarray}}
\newcommand{\bg}{\begin{gather}}
\newcommand{\eg}{\end{gather}}
\newcommand{\bseq}{\begin{subequations}}
\newcommand{\eseq}{\end{subequations}}
\newcommand{\tg}{\mathop{\rm tg}\nolimits}

\renewcommand{\ln}{\mathop{\rm ln}\nolimits}

\begin{document}
\begin{flushright}
CERN-PH-TH/2008-017
\end{flushright}

\vspace*{1cm}
\begin{center}
{\Large \bf Creating semiclassical black holes in collider
  experiments and keeping them on a string}

\medskip
G.~Dvali$^{a,b}$,
S.~Sibiryakov$^{a,c}$
\\
\medskip
$^a${\it Theory Group, Physics Department, CERN, CH-1211 Geneva 23,
  Switzerland.}\\
$^b${\it
Center for Cosmology and Particle Physics, Department of Physics,\\
New York University, New York NY 10003, USA.}\\
$^c${\it
Institute for Nuclear Research of the Russian Academy of Sciences,\\  
60th October Anniversary prospect, 7a, 117312 Moscow, Russia.}
\end{center}
\vspace{0.5cm}

\begin{abstract}

We argue that a simple modification of the TeV scale quantum gravity
scenario allows production of semiclassical black holes in
particle collisions at the LHC. The key idea is that in models with
large extra dimensions the strength of gravity in the bulk can be
higher than on the brane where we live. A well-known example of this
situation is the case of warped extra dimensions. 
Even if the
energy of the collision is not sufficient to create a black hole on the
brane, it may be enough to produce a particle which accelerates 
into the bulk up to trans-Planckian energy and creates a large black
hole there. 
In a concrete model we consider, the black hole is formed in a
collision of the particle 
with its
own image at an orbifold plane.
When the particle in question
carries some Standard Model gauge charges the created black hole gets
attached to our brane by a string of the gauge flux. For a 4-dimensional
observer such system looks as a long-lived charged 
state with the mass continuously
decreasing due to Hawking evaporation of the black hole. This provides
a distinctive signature of black hole formation in our
scenario. 

\end{abstract}

\section{Introduction}
Searches of the physics beyond the Standard Model (SM) 
at the Large Hadron Collider (LHC) 
are predominantly motivated by the Hierarchy 
Problem, the inexplicable stability of the weak scale versus the Planck mass.
One approach to this issue \cite{ArkaniHamed:1998rs}
is that the hierarchy is maintained 
by a low, around a few TeV, value of  the quantum gravity scale $M_*$ 
which is a natural cutoff of the theory. 
Originally, this idea was proposed in the context of theories with
extra dimensions.  
In order to account for the weakness of the gravitational interaction
at large distances the size of extra dimensions in this approach has
to be relatively large. The SM fields are
then assumed to be localized on the 4-dimensional worldvolume of a
3-brane embedded into the higher-dimensional bulk.
It is worth mentioning that, as understood
recently~\cite{dvali,dvali-redi}, the class of   
theories with low scale of quantum gravity is actually wider and
includes generic theories with large number of particle species.

The common collider signature of theories with low quantum gravity
scale would be production of mini black holes (BHs) once the energy of
particle collision exceeds $M_*$ (see, e.g., the second reference in
\cite{ArkaniHamed:1998rs}).  On the other hand the theoretically 
best understood properties, such as the Beckenstein
entropy~\cite{entropy} and Hawking
radiation~\cite{Hawking}, are related to semiclassical BHs
with large masses, $M_{bh}\gg M_*$. It is important to know the 
chances of observing 
such BHs at the LHC. This subject was discussed in
Refs.~\cite{Banks:1999gd,Emparan:2000rs,largebh,
Giddings:2001bu,Dimopoulos:2001hw}.

The standard approach
\cite{Banks:1999gd,Emparan:2000rs,Giddings:2001bu,Dimopoulos:2001hw} 
(see \cite{Giddings:2007nr} for a recent
review)
to BH production at colliders assumes that BHs are
formed directly in the collision of two SM particles.
Then 
the most experimentally 
accessible BHs are the light ones with the mass $M_{bh}$ near the
quantum gravity scale $M_*\sim$ TeV. However, the production rates
and signatures of such BHs are clouded by theoretical
uncertainties.
In contrast to large BHs with $M_{bh}\gg M_*$ which
evaporate into a high multiplicity thermal final state via Hawking
radiation, the light BHs are likely to decay into only a few
high energy particles: a signal difficult to disentangle from the
background. Besides, there are several effects which can push the
collision energy needed for BH formation considerably higher than
$M_*$ \cite{Meade:2007sz}. These include the energy loss by colliding
particles prior to the formation of the BH horizon
and the effects of the charge 
\cite{Yoshino:2006dp}. 
Within the standard scenario this makes
observation of semiclassical BHs at the LHC problematic even in the most
optimistic case $M_*\sim 1$TeV and virtually impossible if 
$M_*\gtrsim 10$TeV.   

In this paper we take another approach.
Generically, theories with large extra dimensions contain particles
which can freely propagate in the bulk. These particles can be
produced in collider experiments once the extra dimensions open up. On
the other hand, the scale of quantum gravity may depend on the 
position in extra
dimensions. In particular, this scale can be much lower in the bulk than on
the brane. Thus, one may imagine sending particles from the brane 
to the region of strong gravity and
producing a large (much larger than the Planck length at that point)
semiclassical BH there. This possibility was first pointed out in
Ref.~\cite{largebh}.

A particular class of theories where the strength of gravitational
interaction grows into the bulk is provided by models with warped
compactification. 
We emphasize that we consider the class of models where 
the parameter $M_*$ entering into the gravitational 
action
\be
\label{Sg}
S_g=-M_*^{d-2}\int d^dx \sqrt{|g|}\,R
\ee
is constant over the whole space-time. In the expression (\ref{Sg})
$d$ is the total number of dimensions and $R$ is the Ricci scalar. 
The parameter $M_*$ sets 
the strength of gravitational
interactions from the point of view of a local bulk observer and so 
this strength is constant. However, 
the situation is different from the point
of view of a 4-dimensional observer who can make measurements only on
the brane. Let us clarify this issue.

Consider a general higher-dimensional metric 
invariant under the 4-dimensional Poincare
group,
\be
\label{metr}
ds^2=A^2(y)\eta_{\mu\nu}dx^\mu dx^\nu+\gamma_{ij}(y)dy^idy^j\;,
\ee
where $\eta_{\mu\nu}$ is the 4-dimensional Minkowski 
metric, and the coordinates
$y^i$ represent extra dimensions. 
The SM brane is located at some point $y=y_b$; we set $A(y_b)=1$.
Generically, the warp factor $A^2(y)$ in front
of the 4-dimensional metric depends on the position in
extra dimensions. In particular, it can be smaller
in the bulk than on the brane, $A(y)<1$ for some $y$. 
Imagine now that two particles are
sent from the brane into the bulk along parallel trajectories
with separation $\Delta x$ between them. Let us neglect for the moment
gravitational attraction between the particles. Then, the coordinate
distance $\Delta x$, which characterizes the separation between the particles
for the brane observer, stays constant while the
distance measured in the local reference frame of 
the bulk observer decreases as
\[
r=A(y)\,\Delta x\;.
\]  
A similar effect occurs for the energies of the particles. While
for the brane observer the energies of the particles remain constant
and equal to their initial values ${\cal E}^{(1)}$, ${\cal E}^{(2)}$,
for the bulk observer the energies are blue-shifted due to the warp
factor,
\[
%\label{blueshift}
E^{(i)}={\cal E}^{(i)}/A(y)\;.
\]
Consider now gravitational interaction between the
particles. The strength of this interaction is characterized by the
dimensionless ratio of the potential gravitational energy $U$ to the
total energy of the system. In the linear approximation the bulk
observer estimates this ratio as,
\[
\frac{-U}{E^{(1)}+E^{(2)}}\sim 
\frac{E^{(1)}E^{(2)}}{M_*^{d-2}r^{d-3}(E^{(1)}+E^{(2)})}\;.
\] 
When expressed in terms of the quantities relevant for the brane
observer the ratio takes the form,
\[
\frac{-U}{E^{(1)}+E^{(2)}}\sim 
\frac{{\cal E}^{(1)}{\cal E}^{(2)}}{[A(y)M_*]^{d-2}\Delta x^{d-3}
({\cal E}^{(1)}+{\cal E}^{(2)})}\;.
\]
We see that the strength of the gravitational interaction of particles
at point $y$ in the bulk, when viewed from the brane, is governed by
an effective Planck mass
\[
M_*^{eff}=A(y)M_*\;.
\]
According to our assumption about the function $A(y)$ this effective
Planck mass decreases from the brane into the bulk. In
this sense we say that gravity in models with warped geometry 
is stronger in the bulk than on the brane.

Clearly, the process consisting of 
simultaneous production of two bulk particles and their
subsequent collision in the strong gravity region to create a BH is
highly improbable. At the same time, from the bulk point of view it is
clear that a single particle cannot produce a BH unless there is some
mechanism to convert its kinetic energy into a center-of-mass energy
of a collision. Luckily, there is a natural candidate for such
mechanism. Models with large extra dimensions generically contain
various objects, such as branes, in the bulk. Collision of the
particle with these objects may well produce a BH. An example which
allows for a particularly elegant description is collision of the
particle with an orbifold fixed point. The latter process can be
represented as collision of the particle with its own image(s) under
the orbifold symmetry (see Figs.~\ref{fig1},~\ref{fig4}). A BH is
expected to be 
produced once all the images (including the particle itself) are
concentrated inside the Schwartzschild radius corresponding to their
total energy. 

In this scenario the BH is created far from the SM brane along the
extra dimensions. So a natural question is how such a BH can be
detected by an observer on the SM brane? We argue that indirect
observation of the BH and its properties, in particular, its Hawking
evaporation, is possible if the particle which creates the BH is
charged under the SM gauge group. The latter situation is not
unrealistic since the bulk particles in theories with extra dimensions
often correspond to the bulk modes of the SM fields. Then, under quite
general assumptions described below, the particle flying away from the
brane will stretch a string of gauge flux between the SM brane and
itself. This string persists when the particle creates BH and connects
the BH to the SM brane. A brane observer perceives the whole
configuration as a charged bound state. Studying the properties of
this bound state she can get information about the BH. This opens a
possibility of (indirect) observation of semiclassical BHs at the LHC
for $M_*$ (the quantum gravity scale on the brane) up to a thousand TeV.

The paper is organized as follows. In the next two sections we
consider a toy model realizing the scenario outlined above. This
enables us to work out the conditions for BH production and the
signatures of this event. Possible generalizations of the model are
discussed in Sec.~\ref{Sec3} which also contains summary of the
results. 
Some technical details are contained in the
appendices.

\section{Toy model}

Let us consider a $d$-dimensional metric\footnote{Our
convention for the signature of the metric is $(+,-,-,\ldots)$.},
\be
\label{RS}
ds^2=A^2(y)\eta_{\mu\nu}dx^\mu dx^\nu
-dy^2-\gamma_{ij}(\theta)d\theta^i d\theta^j\;
\ee
where
\[
%\label{A}
A(y)=\e^{-ky}
\]
and the coordinates $\theta^i$, $i=1,\ldots, d-5$, 
describe a $(d-5)$-dimensional
compact manifold ${\cal M}$ with the metric $\gamma_{ij}$ independent
of $y$ and $x^\mu$. The size $D_{\cal M}$ of the manifold ${\cal M}$
is assumed to be ``large'' in the sense of \cite{ArkaniHamed:1998rs}, 
in particular, we consider the case
$D_{\cal M}\gg k^{-1}\gg M_*^{-1}$. The space-time is cut off by two
codimension-one branes
placed at $y=0$ and $y=y_c$. The positions of these branes are assumed
to be fixed points of $S^1/Z_2$ orbifold. 
We do not pretend to describe a realistic configuration by the metric
(\ref{RS}). Rather, we choose the metric (\ref{RS}) for its simplicity
allowing us
to introduce the main ideas of the paper 
in a transparent way.
In
the case $d=5$ (no $\theta$-dimensions) the setup reduces to the
well-known Randall--Sundrum model \cite{Randall:1999ee}. For $d>5$ 
and ${\cal M}$ a sphere
the
metric (\ref{RS}) is a solution 
of Einstein
equations with a negative cosmological constant and a $(d-6)$-form in
the bulk~\cite{Gherghetta:2000jf}. 

To localize the SM fields we introduce a 3-brane at
$y=\theta_i=0$. 
For simplicity we neglect the back-reaction of this
brane on the metric. This brane will be referred to as
the ``SM brane''\footnote{It is worth stressing 
the difference between our
  approach and that of \cite{Randall:1999ee}. In our setup the SM
  brane is placed at the maximum of the warp factor $A$. This is
  reminiscent of the second Randall--Sundrum model
  \cite{Randall:1999vf}. The hierarchy between $M_*$ and the
  4-dimensional Planck
  scale $M_{Pl}$ is
  achieved due to the large size of the $\theta$-dimensions, see
  Eq.~(\ref{M4})}. 
It is straightforward to find the relation between the parameters of
the model and the four-dimensional Planck mass determining the
strength of gravitational interaction at large
distances on the SM brane. Assuming $y_c\gtrsim k^{-1}$ one obtains,
\be
\label{M4}
M_{Pl}^2\sim M_*^{d-2}V_{\cal M}\,k^{-1}\;,
\ee
where $V_{\cal M}$ is the volume of the manifold ${\cal M}$. 
By the standard arguments 
one concludes that $M_*$ can be as low as a few TeV if the  product
$V_{\cal M}\,k^{-1}$ is large enough.

Apart from the SM matter localized on the brane, models with large
extra dimensions generically contain matter propagating in the
bulk. For example, 
presence of bulk matter is unavoidable in any field-theoretical
realization of the localization mechanism: it corresponds to the bulk
modes of the fields whose localized excitations describe the SM
particles. The rate of production of the bulk modes in
collisions of SM particles is model dependent. Typically, it becomes
significant 
above a certain
threshold energy $E_{th}$. The existing experiments set the bound 
$E_{th}\gtrsim 1$TeV.

\begin{figure}[htb]
\begin{center}
\includegraphics[width=0.4\textwidth]{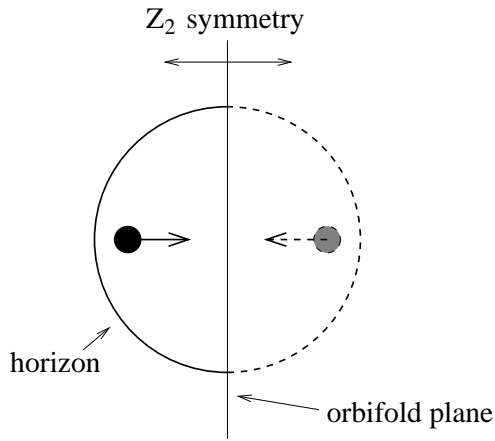}
\end{center}
\caption{Collision of the particle with an orbifold plane viewed as 
collision of the particle with its image under $Z_2$
symmetry. The black hole horizon is shown schematically.}
\label{fig1}
\end{figure}

Consider a bulk particle produced on the SM brane with energy
$E_0\gtrsim E_{th}$. To make the argument as clear as possible 
let us assume for the moment that the 
particle moves along the $y$-direction. 
The time independence of the metric
(\ref{RS}) implies conservation of the ``coordinate'' energy ${\cal E}$. The
latter is related to the physical energy $E$ of the particle in 
the local
inertial frame as follows,
\[
%\label{EE}
{\cal E}=A(y)E\;.
\] 
Thus, the particle reaches the plane $y=y_c$ with the physical energy 
$
%\label{Ec}
E_c=E_{0}/A_c
$,
where $A_c\equiv A(y_c)$. 
Here we arrive to an important point in our discussion. The
requirement of $Z_2$ symmetry of the metric across the plane $y=y_c$
implies that the gravitational field of a particle approaching this
plane is equal to the gravitational field of the pair consisting of
the particle and its image under $Z_2$ reflection, see
Fig.~\ref{fig1}. 
In other words,
collision of the particle with the orbifold plane is equivalent to the
head-on collision of the particle with its own image. If $E_c>M_*$
this collision results in BH formation 
\cite{D'Eath:1992hb,Eardley:2002re}. The mass of the BH is estimated
as 
\[
%\label{Mbh}
M_{bh}\sim E_c=E_{0}/A_c\;.
\] 
The semiclassical description is valid
when $M_{bh}\gg M_*$. 

In a curved background, such as ours, the properties of a BH are
different in the cases when its size is smaller or larger than the
curvature radius of the background. Small BHs are not affected by the
background curvature and are essentially the same as BHs in flat
space-time. Their horizon radius is given in $d$ space-time dimensions
by the formula  
\[
%\label{Rh}
R_h\sim \big(M_{bh}/M_*^{(d-2)}\big)^{1/(d-3)}\;.
\] 
On the other hand, large BHs with $R_h>k^{-1}$ are sensitive to the
characteristics of the background. An exact solution describing
a large BH in the background (\ref{RS}) is unknown; it is not even
clear if a stable solution exists at all. 
The analysis of the Randall--Sundrum case (no large
$\theta$-dimensions) performed in \cite{Giddings:2002cd} shows that
the form of the solution, if any, 
crucially depends on the properties of the
mechanism stabilizing the interbrane distance. To avoid these
complications we consider only small BHs with
$R_h\lesssim k^{-1}$ in what follows. 
This constrains the allowed BH mass from above,
\be
\label{bhflat}
M_{bh}\lesssim M_* (M_*/k)^{d-3}\;.
\ee 
As $k$ is much smaller than $M_*$ this requirement is not very
restrictive and
leaves a wide parameter region for BH production.

The above discussion is easily generalized to the case when the
particle traveling in the bulk has non-zero momentum along dimensions
other than the $y$-direction. All the formulae remain valid with the
substitution of the total energy $E$ by the energy of the motion in
the direction perpendicular to the orbifold plane,
\[
%\label{Eperp}
E_{\perp}=\sqrt{E^2-p_\parallel^2}\;,
\] 
where $p_\parallel$ is the component of the particle momentum
along the SM brane. Note that the motion of the particle along the 
$\theta$-directions can be neglected as the corresponding components
of the particle momentum are not blue-shifted and hence
get small compared to the total energy   
by the
moment the particle reaches the orbifold plane.
Evidently, the BH produced in this general case is not at rest: 
it slides along
the orbifold plane. 

Thus we have identified an efficient mechanism of BH
production. Particles leaving the SM brane get accelerated in the bulk
and create BHs in collisions with the distant orbifold plane. Note
that this process is possible for any value of $M_*$ provided the orbifold
plane is far enough. We now turn to the question under what conditions
creation of BH can be detected by an observer on the SM brane. 
The BH emits Hawking radiation with temperature
\be
\label{TH}
T_H\sim \big(M_*^{d-2}/M_{bh}\big)^{1/(d-3)}\;.
\ee
The temperature gradually increases as the BH evaporates. 
The problem is that the radiation proceeds into the bulk modes which are
hard to detect on the brane.
Indeed, the temperature $T_0$ of the
radiation when it reaches the brane is red-shifted compared to
(\ref{TH}),
\be
\label{T01}
T_0=T_H A_c\;.
\ee
The bulk modes interact appreciably with the SM matter only if their
energy 
exceeds the threshold for brane--bulk interaction, 
$T_0\geq E_{th}$.
Thus, the radiation of the BH becomes visible on the brane only
during the late stages of BH evaporation corresponding to 
$T_H\geq E_{th}A_c^{-1}$. This temperature should be still smaller
than the quantum gravity scale,
$T_H<M_*$. Combined with the condition for the BH production this
implies the following inequalities,
\be
\label{ineq}
E_{th}A_c^{-1}<M_*<E_0A_c^{-1}\;.
\ee
In principle, the requirement (\ref{ineq}) leaves non-empty range of
parameters. However, in
practice $E_0$ cannot be much larger than $E_{th}$. This restricts the
masses of the produced BHs to be near the quantum gravity scale $M_*$
and thus brings back the theoretical uncertainties discussed in the
Introduction. 
These preliminary considerations show that this scenario of BH
detection,
though not completely ruled out,
is quite
problematic.
Note that in the case of more than one large
codimension ($d>5$) the scenario suffers from additional
complications due to the dilution of BH signal in large dimensions
transverse to the SM brane. In what follows we concentrate on another
option.

\section{String of gauge flux}

\begin{figure}[htb]
\begin{center}
\includegraphics[width=0.35\textwidth]{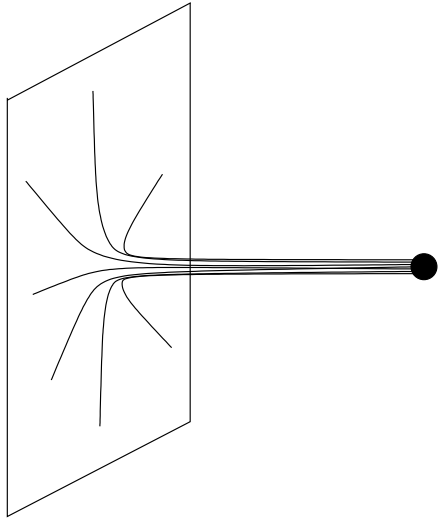}
\end{center}
\caption{Charged particle placed into the bulk is
  connected to the brane by a string of gauge flux. The flux opens
  onto 
  the brane as the field of a point charge.}
\label{fig2}
\end{figure}

A theory
with large extra dimensions implies a mechanism of localization of
the SM gauge fields on the brane. One generic mechanism  
\cite{Dvali:1996xe} (see also \cite{Dubovsky:2001pe}) assumes that the
gauge theory is in the confining phase in the bulk and thus possesses
a mass gap $\mu$. On the brane confinement switches off and the gauge
fields become massless. They cannot fly away from the brane if their
energy is smaller than $\mu$, i.e. they are 
localized.
The experimental limit on the scale of the bulk
confinement is $\mu\gtrsim 1$TeV. Assume now that there exist bulk
particles charged under 
the SM gauge
group. This is true, for example, if these particle are the bulk
modes of the SM fields. When
such a particle leaves the SM brane after being created in a high
energy collision its gauge flux concentrates into a string stretching
between 
the particle and the brane, see   
Fig.~\ref{fig2}. This string ensures conservation of gauge charges
from the point of view of the observer on the brane who perceives the
end of the string as a point charge.

Let us make a digression and discuss the following issue. An explicit
realization of the mechanism of Ref.~\cite{Dvali:1996xe} for
non-Abelian fields in more than four space-time dimensions is
unknown. However, the inevitability of the flux-tubes (open strings)
attached to the brane goes beyond the specifics of the mechanism
of Ref.~\cite{Dvali:1996xe}. It follows from just two assumptions: dynamical
localization of a massless gauge field on the brane and existence of a
mass gap in gauge theory in the bulk. 
Indeed, one can imagine performing a thought experiment in which one
removes the test charge away from the brane.  Because of the
conservation of the gauge charge on the brane, guaranteed by the
existence of the zero mode photon and the bulk mass gap, 
an observer on the brane should
measure the same gauge flux through any two-sphere on
the brane that surrounds the point from which the charge was removed
to the bulk. Because the flux lines cannot break, the only option is
that the removed charge continues to produce exactly the same amount
of the gauge flux in form of a tube that terminates on the 
brane\footnote{Note that universality of the flux tubes is, in particular,
supported by the properties of D-branes in string theory 
where the presence
of massless gauge fields on the brane is related to existence of open
strings attached to it.}.
It is worth emphasizing the importance of the bulk mass gap in the
above reasoning: in the absence of the mass gap the situation is more
involved~\cite{Dubovsky:2000av}.

We now return to the consequences of the existence of gauge flux strings
for our BH production scenario. Let the particle which produces
the BH carry some of the SM gauge charges. The same charges will be
carried by the resulting BH, and therefore the latter will be
connected to the SM brane by the string. 
Probing the end of the string on the SM brane one may extract 
information about the BH.
We proceed to a more careful analysis of this
possibility. 

Let us specify the acceptable range of parameters. First, we consider
the case 
\be
\label{muk}
\mu>k\;,
\ee
so that the effects of the space-time curvature on the inner structure
of the string can be neglected. 
Second, one has to check that the confining force of the string does not
prevent the energetic particle from leaving the SM brane and producing
a BH.
To this end let us  
calculate the conserved energy of the string 
stretching between the
SM brane and
the orbifold plane at $y=y_c$. 
Taking the string tension to be equal to $\mu^2$
one obtains
\be
\label{energy}
{\cal E}_{str}=\int_0^{y_c}\mu^2 A(y)dy=
\frac{\mu^2}{k}\left(1-\e^{-ky_c}\right)\;.
\ee
We see that the energy of the string, as seen by
the SM observer, is effectively concentrated in the piece of length
$k^{-1}$ near the brane and  
approaches the asymptotic
value $\mu^2/k$ at $y_c\gg k^{-1}$. 
The gravitational force pulling 
the
particle into the bulk overcomes the confining force
of the string if the energy of the
particle is large enough,
\be
\label{Emu}
E_{0}>\mu^2/k\;.
\ee
As a side remark let us note that particles with energies less
than $\mu^2/k$ are confined to the SM brane in spite of the gravitational
pull. Therefore the gauge field localization mechanism considered here
provides also localization of light charged
fields.

\begin{figure}[htb]
\begin{center}
\includegraphics[width=0.8\textwidth]{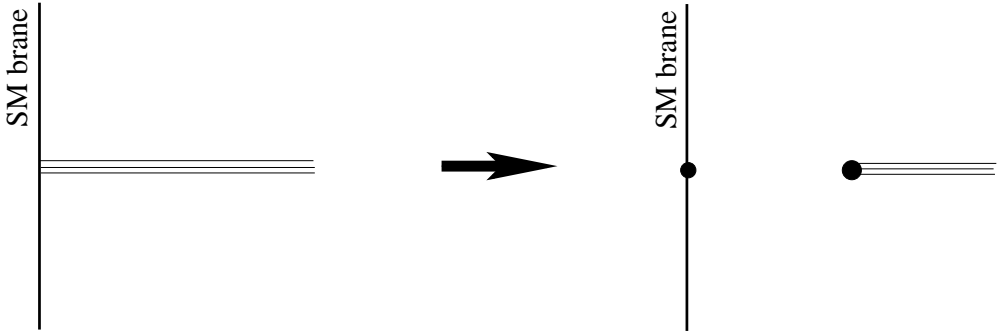}
\end{center}
\caption{The string can break off the brane by producing one
  heavy bulk particle and one light particle left on the brane.}
\label{fig5}
\end{figure}

One more restriction on parameters comes from the requirement that the
string should be stable on the time-scales involved in the problem.   
The physical mechanism which destabilizes the string is
pair production
of charged particles analogous to the Schwinger process. As shown in
Appendix~\ref{app:A}, the fastest instability corresponds to the string
breaking off the brane by production of one bulk particle and
one light SM particle, see Fig.~\ref{fig5}.
If the
mass $m_{ch}$ of the lightest charged particle in the bulk is larger
than $\mu$
this process is exponentially suppressed. The width of this process is
estimated in Appendix~\ref{app:A},
\be
\label{Gam}
\Gamma_{str}\sim\mu\exp\left[-\frac{2 m^2_{ch}}{\mu^2} 
~f\left(\frac{m_{ch}k}{\mu^2}\right)\right]\;,
\ee
where the function $f(x)$ has the form, 
\be
\label{ff}
f(x)=\frac{\pi-2x+2\sqrt{x^2-1}\,\ln(x+\sqrt{x^2-1})}{2x^2}\;.
\ee
Below we will see that the string instability 
can be slow enough to allow detection of BHs.

Now we are ready to work out the signal produced on the SM brane by
the BH attached to the brane by the string.
At the moment of
production the BH--string configuration looks for an observer on the
brane as a bound state with the 
mass\footnote{For the sake of the argument we assume that the
  configuration is produced almost at rest and 
neglect energy loss in the process of
BH formation.} 
$E_0$. The bound state consists of a charged particle (corresponding
to the string in the higher-dimensional language) with mass 
$\mu^2/k$ and a neutral component (corresponding to BH). 
As the BH evaporates the mass of the neutral component, and hence of
the whole bound state, decreases.
This process continues until the Hawking temperature of the
BH reaches the value $m_{ch}$. Using the expression (\ref{TH}) for the
BH temperature one obtains that this corresponds to the BH mass
\be
\label{M1}
M^{(1)}_{bh}\sim M_*(M_*/m_{ch})^{d-3}\;.
\ee   
The
mass of the bound state observed on the SM brane at this
moment is
\[
%\label{E1}
E_1\approx \frac{\mu^2}{k}+
A_cM_*\left(\frac{M_*}{m_{ch}}\right)^{d-3}\;.
\]
At $T_H>m_{ch}$ it becomes possible for the BH to separate from the
string by emitting a charged bulk particle. 
After that the BH
and the string exist independently of each other: the BH has slipped
off the hook.
For the brane observer
this corresponds to the splitting of the bound state into charged and
neutral pieces. Only the charged component with the 
mass\footnote{The small contribution
$A_cm_{ch}$ to the total mass of the charged component, coming from the 
particle terminating the string on the orbifold plane, can be
neglected.}  
${\cal E}_{str}$, see Eq.~(\ref{energy}),
continues to be visible to the brane observer.

Let us estimate the lifetime $\Delta t$ of the BH--string bound state. 
One writes down
\[
\frac{dM_{bh}}{dt_c}\sim -T^2_H\;,
\]
where $t_c=A_ct$ is the physical time at the BH position. Using
Eq.~(\ref{TH}) and integrating over the BH masses from the 
value at production, $M_{bh}^{(0)}\approx E_0/A_c$, to the value (\ref{M1}) at
the break-off of the bound state one obtains
\[
%\label{Dt}
\Delta t\approx\frac{1}{A_cM_*}
\left[\left(\frac{M_{bh}^{(0)}}{M_*}
\right)^{\frac{d-1}{d-3}}
-\left(\frac{M_*}{m_{ch}}\right)^{d-1}\right]\;.
\]
Clearly, positivity of $\Delta t$ requires
$M_{bh}^{(0)}$ to be larger than $M_{bh}^{(1)}$, Eq.~(\ref{M1}). 
The lifetime $\Delta t$ of the bound state can be very large, see
Eq.~(\ref{Deltat}) below. It makes the signature of BH production very
distinctive:
formation of a heavy charged long-lived 
bound state with the mass slowly changing
in time.

Further information can be obtained by monitoring the charged
component left after the splitting of the BH--string bound state. The
behavior of this charged component  
depends on the ratio between $m_{ch}$ and 
$\mu^2/k$. If $m_{ch}>\mu^2/k$ the gravitational pull forces the 
bulk particle terminating the string to remain on the orbifold plane.
The string remains stretched between the orbifold plane and the SM
brane until it breaks away from the brane after
time $\Gamma_{str}^{-1}$.
What is seen 
by the brane observer is
a heavy charged particle with mass ${\cal E}_{str}$ 
decaying with lifetime
$\Gamma_{str}^{-1}$ into a SM particle with the same charges plus
something invisible. Conversely, if $m_{ch}<\mu^2/k$ the
string collapses onto the SM brane immediately after the separation
from the BH and releases its energy 
decaying into the SM fields.

The above discussion is valid if the lifetime 
of the BH--string bound state
is smaller than the decay time of the string,
\be
\label{DtGam}
\Delta t<\Gamma_{str}^{-1}\;.
\ee
Another requirement is that the BH--string system must be bound
strongly enough to be considered as a single object. This amounts to
requiring that the frequencies of internal oscillations of the bound
state 
are large compared to its inverse lifetime. In Appendix~\ref{app:B} we
find that the spectrum of the internal oscillations of the BH--string
system consists of 
of a low
lying mode with frequency
\be
\label{omega0}
\omega_0=A_c^{3/2}k\sqrt{3\left(1+\frac{\mu^2}{A_ckM_{bh}}\right)}
\ee
and a tower of modes with frequencies,
\be
\label{omegan}
\omega_n\approx \pi A_ck(n+1/2)~,~~~~~n=1,2,\ldots\;.
\ee 
Note that the lowest frequency (\ref{omega0}) is sensitive to the BH
mass: it grows as the BH evaporates. On the other hand the higher
eigenfrequencies (\ref{omegan}) correspond to the intrinsic
oscillations of the string: they do not depend on the mass of the BH
and are present even after the string separates from the BH. The
BH--string system is strongly bound if
\be
\label{omegaDt}
\omega_0\gg \Delta t^{-1}\;.
\ee 

The quantities entering into Eqs.~(\ref{DtGam}), (\ref{omegaDt})
have complicated dependence on the parameters of the model.
A careful study is needed to work out the 
region of parameter space where these requirement
as well as other
constraints (\ref{bhflat}), (\ref{muk}), (\ref{Emu}) 
are satisfied. Such a study is beyond the scope of the present
article. We limit ourselves to rough estimates which show that the
allowed 
region is likely to contain
phenomenologically interesting values. As an example we consider  
\be
\label{param}
\begin{split}
&d=7~,~~~M_*=1000\mathrm{TeV}~,~~~A_c=10^{-12}\;,\\
&k=0.2\mathrm{TeV}~,~~~\mu=1\mathrm{TeV}~,~~~E_0=m_{ch}=8\mathrm{TeV}\;.
\end{split}
\ee
Using the equations given above
one obtains,
\begin{align}
&M_{bh}^{(0)}=3\cdot 10^{15}\mathrm{GeV}~,~~~~~
M_{bh}^{(1)}=2.4\cdot10^{14}\mathrm{GeV}\;,\notag\\
&E_1=5.24\mathrm{TeV}~,~~~~~{\cal E}_{str}=5\mathrm{TeV}\;,\notag\\
&\Delta t\approx 1.6\cdot 10^{20}\mathrm{GeV}^{-1}
\approx 10^{-4}\mathrm{s}\;,
\label{Deltat}\\
&\Gamma_{str}^{-1}\sim 0.6\cdot 10^{25}\mathrm{GeV}^{-1}\approx 
4\mathrm{s}\;,\notag\\
&\omega_0=(5.7\div 16)\cdot 10^{-16}\mathrm{GeV}\approx
(0.8\div 2.4)\cdot 10^9\mathrm{s}^{-1}\;,\notag
\end{align}
where in the last line we have indicated the range of variation of
$\omega_0$ due to the change in the BH mass.
It is straightforward to
check that these values indeed satisfy all
the constraints.
Note that the lifetime of the BH--string bound state and the decay time
of the string  are very large by 
particle physics standards. 

Let us conclude the analysis of our toy model by indicating a few more 
possible signatures of the BH--string bound system. As mentioned
above the lowest frequency of this system depends on time due to the
change in the BH mass. A measurement of this dependence would
give an accurate information on the BH evaporation process. Such a
measurement 
could be particularly useful in the case of a very slow BH evaporation
implying a  
long lifetime of the
BH--string
bound state. 
We have
nothing to say about the experimental feasibility of such a
measurement. 

Finally, it is known \cite{Frolov:2000kx} 
that a string embedded into a BH 
provides an additional channel for the Hawking 
radiation. This effect gives rise to thermal
oscillations of the string in the BH--string system. 
The corresponding temperature $T_0$, as
seen from the SM brane, is given by Eq.~(\ref{T01}). The typical value
of $T_0$ is quite small compared to the mass of the constituents of the
BH--string 
system. For
example, the choice (\ref{param}) gives $T_0\sim 1$eV. It is not
clear if these oscillations can lead to
observable effects on the brane. We leave this issue for future studies.

\section{Generalizations and summary}
\label{Sec3}

\begin{figure}[htb]
\begin{center}
\includegraphics[width=0.4\textwidth]{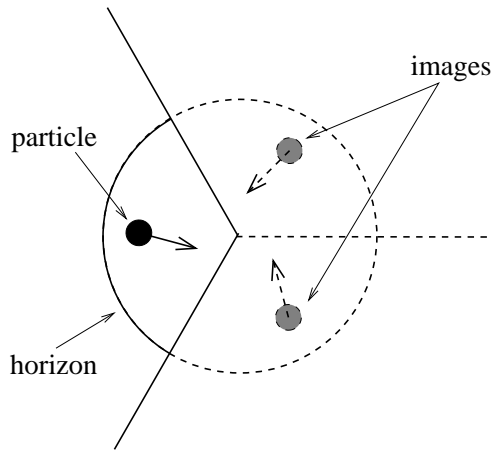}
\end{center}
\caption{Collision of particle with an $\mathbb{R}^2/Z_3$ orbifold
  fixed manifold is equivalent to collision with two images. The
  collision produces a black hole.}
\label{fig4}
\end{figure}

The toy model studied above admits various
generalizations. With slight changes of the formulae of the previous
sections the scenario is applicable to a wide class of higher-dimensional
backgrounds. 
One class of generalizations concern the mechanism
of BH production. Instead of considering collision of a
particle with an orbifold plane
one may consider 
collision with an orbifold fixed manifold
of codimension higher than one. Such a collision also
results in BH formation, see Fig.~\ref{fig4} where a collision of the particle
with an ``orbifold string'' is shown as an example. 
More generally,
BHs may result from particle collisions with other branes present in
the bulk. 

Other generalizations correspond to the case when the parameters
governing the strength of gravitational and gauge interactions vary
across the bulk. For example, gravity can be coupled  
to a dilaton field
so that the higher-dimensional gravitational action takes the form,
\[
S_g=-M_*^{d-2}\int d^dx~\sqrt{|g|}\,R\,\e^{-\phi}\;.
\] 
In this case the strength of gravitational interaction depends on the
position in the extra dimensions even from the point
of view of bulk observers: 
if $\phi$ has a non-trivial
profile in the extra dimensions the local Planck scale
$M_*\e^{-\phi/(d-2)}$ varies accordingly. In particular,
gravity may
become stronger deep in the bulk than near the SM brane. Yet another
possibility is that the gravitational interaction on the
brane is suppressed compared to the bulk by
the presence of a brane 
Einstein--Hilbert term in the gravitational 
action~\cite{largebh,Dvali:2000hr}. All these effects, together with
possible dependence of 
the confinement scale $\mu$,
the mass of the charged particle $m_{ch}$ and the background curvature
scale $k$ on the position in the bulk, can affect 
the properties (mass, lifetime, etc.) of the
string and of the BH--string bound state. 

To sum up, 
we have proposed a generic scenario of production and
detection of semiclassical BHs in the framework of theories with large
extra dimensions. In contrast to the standard picture of BH
production in particle collisions the BHs in our scenario
are created away from the
brane in the regions of the bulk where gravity is stronger than on the
brane. The signatures of the scenario are largely different from those
usually studied in the context of BH production.

The details of a given
realization of this scenario are model dependent. However, the main
ideas are summarized as follows.  
A collision on the SM brane produces a particle which can freely
propagate in the bulk. This particle flies away from the brane into
the region where its energy is much higher than the local Planck
scale.
There it produces a BHs in collisions with some
object (orbifold
fixed manifold, brane, etc.) in the bulk. 
If the bulk particle is charged under
the SM gauge group, the
resulting BH is connected to the SM brane by a string of gauge flux. 
From the point
of view of the SM observer the system behaves as a heavy bound state
with charged and neutral components. The mass of the latter component
slowly decreases with time due to evaporation of the BH. After some, 
generically large, time the BH separates from the string 
which corresponds to the break off of
the bound state.
This behavior provides
a distinctive signature of our scenario.

Other possible signatures
depending of the details of a given realization
include observation of low-frequency excitations
of the bound state, thermal oscillations of the system due to BH
Hawking radiation, or decay of the charged component after the
break off of the bound state. It would be interesting to 
study the signatures of the scenario proposed in this paper in more
detail and construct viable extra-dimensional models where this
scenario is realized.

{\bf Acknowledgments.} We thank D.~Gorbunov and M.~Redi 
for useful
discussions. We are grateful to
O.~Pujolas and I.~Tkachev
for encouraging interest. This work was supported in part 
by the David and Lucile Packard Foundation Fellowship for Science and
Engineering, by NSF grant PHY-0245068 and by 
EU 6th Framework Marie Curie
Research and Training network "UniverseNet" (MRTN-CT-2006-035863).

\appendix
\section{Stability of the string in warped background}
\label{app:A}

In this appendix we study the process of string breaking by pair
production of charged particles in the warped background
(\ref{RS}). We consider a string stretched between the SM brane at
$y=0$, $\theta_i=0$ and the point $y=y_c$, $\theta_i=0$ on the
orbifold plane. If the mass $m_{ch}$ of the charged particles in the
bulk is larger than the scale $\mu$ of the string tension, the pair
production is a tunneling process and can be analyzed by semiclassical
methods. 

The case of an infinite string in flat space-time was studied in
\cite{Vilenkin:1982hm} where the probability of string decay per unit
time per unit length was found,
\be
\label{prob}
P=\mu^2 A\left(\frac{m_{ch}}{\mu}\right)
\exp\left[-\frac{\pi m_{ch}^2}{\mu^2}\right]\;,
\ee
where $A(x)$ is some function whose form is not important as long as
the exponential factor is small. We will omit this function in what
follows. The result (\ref{prob}) is also applicable in curved
backgrounds if the distance $R=m_{ch}/\mu^2$ 
between the particles at the moment of
production is small compared to the curvature radius of the
background. In our case this corresponds to 
$m_{ch}k/\mu^2\ll 1$.
Integrating (\ref{prob}) over the length of the string and taking into
account the dilation of the time intervals due to warping we obtain
the width of the string decay via pair production in the bulk for the
case $m_{ch}k/\mu^2\ll 1$,
\be
\label{Gam1}
\Gamma_{bulk}=\int PA(y) dy\sim\frac{\mu^2}{k}
\exp\left[-\frac{\pi m_{ch}^2}{\mu^2}\right]\;.
\ee
On the other hand, the pair production in the bulk is completely
suppressed by gravitational effects if $m_{ch}k/\mu^2>1$. Indeed,
consider the change in the conserved energy of the system when the
string breaks into two parts terminated by charged particles
at the positions $y=y_1$, $y=y_2$. One obtains,
\[
%\label{DE}
\Delta {\cal E}=\left(m_{ch}-\frac{\mu^2}{k}\right)\e^{-ky_1}
+\left(m_{ch}+\frac{\mu^2}{k}\right)\e^{-ky_2}\;.
\] 
We see that if $m_{ch}k/\mu^2>1$ the change of energy is positive and thus
the pair production in 
the bulk is
impossible.

In our case there is one more channel of the string instability
compared to the case of the infinite string. Namely, the string can
break away from the brane via pair production of a bulk particle
and a SM particle, see Fig.~\ref{fig5}. 
As this process requires
production of only one heavy bulk particle one may expect that it is
preferable to the bulk pair production considered above. We are going
to see that this is indeed the 
case.

\begin{figure}[htb]
\begin{center}
\includegraphics[width=0.25\textwidth]{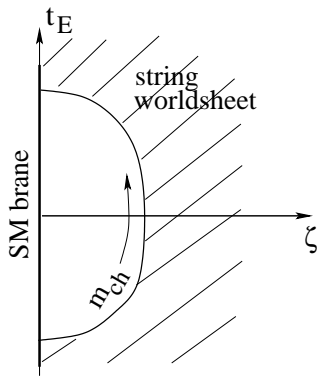}
\end{center}
\caption{The Euclidean solution describing the string breaking away
  from the brane.}
\label{fig6}
\end{figure}

On general grounds the width of the process depicted in
Fig.~\ref{fig5} has the form
\be
\label{Gam2}
\Gamma=\mu B(m_{ch}/\mu,k/\mu)\exp[-S_B]\;,
\ee
where $S_B$ is the action of the bounce solution describing the string
decay, and $B$ is some function which we will omit in what
follows. The bounce solution extremizes the Euclidean action of the
system consisting of the charged particle attached to the end of the
string,
\be
\label{SE}
S_E=\mu^2\int\sqrt{\det\bar g_{ab}}~d^2\xi
+m_{ch}\int ds\;,
\ee
where $\bar g_{ab}$ is the induced metric on the string
worldsheet parameterized by the coordinates $\xi^{a}$,
$a=1,2$; in the second term integration is performed over the
trajectory of the end of the string. In writing down the action
(\ref{SE}) we have neglected the contribution of the light SM particle
living on the brane. The bounce solution has the form of a hole in the
string worldsheet bounded on one side by the trajectory of the bulk
particle and on the other side by the brane, see Fig.~\ref{fig6}.
It is convenient to bring the background metric into the conformally
flat form by introducing the coordinate
\be
\label{zeta}
\zeta=k^{-1}\e^{ky}
\ee 
and parameterize the particle trajectory by $\zeta(t_E)$, where $t_E$
is the Euclidean time. Then the bounce action takes the following
form,
\be
\label{SB2}
S_B=\int dt_E\left[-\frac{\mu^2}{k}+\frac{\mu^2}{k^2\zeta}
+\frac{m_{ch}}{k\zeta}\sqrt{1+\left(\frac{d\zeta}{dt_E}\right)^2}
\right]\;,
\ee
where we subtracted the action of the ``vacuum'' configuration
corresponding to the string without holes. 
Extremization of the action (\ref{SB2}) over
the trajectories of the particle yields
\be
\label{SB3}
S_B=\frac{2m_{ch}^2}{\mu^2}
\int_0^1 dz\frac{\sqrt{1-z^2}}{1+\frac{m_{ch}k}{\mu^2}z}
=\frac{2m_{ch}^2}{\mu^2}~f\left(\frac{m_{ch}k}{\mu^2}\right)\;,
\ee
where the function $f$ is defined in Eq.~(\ref{ff}). Substituting this
expression into Eq.~(\ref{Gam2}) one obtains the formula (\ref{Gam})
of the main text. Note that at $k=0$ equation (\ref{SB3}) gives 
$S_B=\pi m^2_{ch}/2\mu^2$ which is half of the suppression exponent in
Eq.~(\ref{Gam1}). Thus the width of the process of Fig.~\ref{fig5}
is much larger than the width (\ref{Gam1}) of pair production in the bulk.

\section{Internal oscillations of BH--string bound system}
\label{app:B}

Let us study the spectrum of excitations of the BH--string bound
system. 
Of
main interest to us are oscillations of the string in the directions
parallel to the SM brane: these oscillations correspond to the motion
of the charged component of the bound state visible by the brane
observer. 
In what follows we concentrate on this type of excitations and
leave aside possible oscillations of the string in the
$\theta$-directions. 

The worldsheet of the string is described by
the functions $x^\alpha(t,\zeta)$, where $x^\alpha$, $\alpha=1,2,3$
are the spatial coordinates parallel to the brane and 
$\zeta$ is the coordinate along the fifth dimension 
defined in (\ref{zeta}).
To
simplify notations we omit the index $\alpha$ below.
The string is stretched between the SM brane
at $\zeta=k^{-1}$ and the BH sitting at $\zeta_c=k^{-1}/A_c$.
For our present
purposes the BH can be treated as a point mass $M_{bh}$ moving 
along the $x$-direction,
$x_{bh}(t)=x(t,\zeta_c)$. Expanding the Nambu--Goto action of the
string and the action of the point mass around the static
configuration $x(t,\zeta)=0$ one writes down the action
for the small oscillations of the BH--string system,
\[
%\label{strfluct}
S=\frac{\mu^2}{2}
\int \frac{dt d\zeta}{(k\zeta)^2}\;[(\d_tx)^2-(\d_\zeta x)^2]
+\frac{M_{bh}}{2}\int \frac{dt}{k\zeta_c}(\d_t x)^2\bigg|_{\zeta=\zeta_c}\;.
\]
From this action one derives the equation for the string oscillations,
\be
\label{eq}
-\frac{\d^2 x}{\d t^2}+\frac{\d^2 x}{\d\zeta^2}-
\frac{2}{\zeta}\frac{\d x}{\d\zeta}=0\;,
\ee
supplemented by the boundary conditions
\bseq
\label{bc}
\begin{align}
\label{bc1}
&\d_\zeta x\big|_{\zeta=k^{-1}}=0;,\\
\label{bc2}
&\mu^2\d_\zeta x+M_{bh}(k\zeta_c)\d_t^2x\big|_{\zeta=\zeta_c}=0\;.
\end{align}
\eseq
The solutions of Eqs.~(\ref{eq}), (\ref{bc}) have the form
\[
%\label{solut}
x_\omega=C_\omega\left[(1+i\omega\zeta)\e^{-i\omega(\zeta-k^{-1})}
-(1-i\omega\zeta)\e^{i\omega(\zeta-k^{-1})}\right]\e^{-i\omega t}\;,
\]
where $C_\omega$ is an arbitrary constant and
the eigenfrequencies $\omega$ satisfy the following equation,
\[
%\label{freq}
\frac{\tg[\omega(\zeta_c-k^{-1})]}{\omega\zeta_c}=
\left(1-\frac{\mu^2}{kM_{bh}}\right)^{-1}\;.
\]
For $M_{bh}\gg\mu^2/k$ this equation leads to the expressions
(\ref{omega0}), (\ref{omegan}).

\end{document}